\documentclass[12pt]{iopart}
\usepackage{iopams}  
\usepackage{epsf}  
\usepackage{float}  
\begin{document}

\title[Directed Percolation and the Golden Ratio]{Directed Percolation and the Golden Ratio}

\author{Stephan M Dammer\dag\footnote[2]{To whom correspondence 
should be addressed ({\tt dammer@comphys.uni-duisburg.de})}
, Silvio R Dahmen\S \\and
Haye Hinrichsen$\|$}

\address{\dag\ Theoretische Physik, Fachbereich 10,
        Gerhard-Mercator-Universit{\"a}t Duisburg,
        47048 Duisburg, Germany}

\address{\S\ Instituto de Fisica, Universidade Federal do Rio Grande do Sul (UFRGS), 91501-970 Porto Alegre RS Brazil}

\address{$\|$\ Theoretische Physik, Fachbereich 8, Universit{\"a}t
        Wuppertal, 42097 Wuppertal, Germany}

\begin{abstract}
Applying the theory of Yang-Lee zeros to nonequilibrium critical
phenomena, we investigate the properties of a directed bond percolation
process for a complex percolation parameter $p$. It is shown that for the
Golden Ratio $p=(1\pm\sqrt{5})/2$ and for $p=2$ the survival probability of a cluster can be computed exactly.
\end{abstract}

%
%
\nosections
Directed percolation (DP) is an anisotropic variant of ordinary
percolation in which activity can only percolate along a given
direction in space. Regarding this direction as a temporal
degree of freedom, DP can be interpreted as a dynamical process.
Directed percolation represents one of the most prominent universality
classes of nonequilibrium phase transitions from a fluctuating active phase into a
non-fluctuating absorbing state~\cite{Kinzel,MarroDickman,Hinrichsen1,Hinrichsen2}. 

\begin{figure}
\epsfxsize=74mm
\centerline{\epsffile{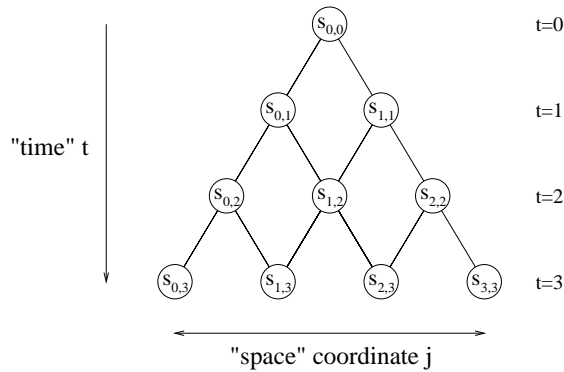}}
\caption{\label{FigBondDP}
Lattice geometry of directed bond percolation in 1+1 dimensions. Sites can be
either active ($s_{j,t}=1$) or inactive ($s_{j,t}=0$). Activity can percolate
forward in time through bonds (solid lines) which are conducting with
probability $p$ and non-conducting with probability $1-p$.}
\end{figure}

A simple realization of DP is directed bond percolation. In this model the
bonds of a tilted square lattice are conducting with probability $p$ and
non-conducting with probability $1-p$ (see Figure~\ref{FigBondDP}). The order
parameter which characterizes the phase transition is the probability
$P(\infty)$ that a randomly chosen site belongs to an infinite cluster. A
cluster consists of all sites that are connected by a directed path of
conducting bonds to the sites that generate the
cluster at time $t=0$. For $p>p_{\rm
  c}$ this probability is finite whereas it vanishes for $p \leq
p_{\rm c}$. Close to the phase transition $P(\infty)$ is known to vanish algebraically as $P(\infty) \sim (p-p_{\rm
  c})^\beta$. Although DP can be defined and simulated easily, it is one of
the very few systems for which -- even in one spatial dimension -- no
analytical solution is known, suggesting that DP is a non-integrable
process. In fact, the values of the percolation threshold and the critical
exponents are not simple numerical fractions but seem to be irrational
instead. Currently the best estimates for directed bond percolation in 1+1
dimensions are $p_{\rm c}=0.6447001(1)$ and $\beta=0.27649(4)$
\cite{Jensen}.

In recent years a large variety of {\it equilibrium} phase transitions have
been analyzed by studying the distribution of Yang-Lee zeros
\cite{YangLee,YangLeeNeuer}. Here the partition sum of a finite equilibrium
system is expressed as a polynomial of the control parameter (usually a
function of temperature). Typically the zeros of this polynomial lie on simpler
geometric manifolds such as circles in the complex plane. As the system size
increases these zeros approach the real axes at the phase transition
point. This explains the crossover to a non-analytic behaviour at the
transition in the limit of an infinite system. Even more recently, the
concept of Yang and Lee has been generalized to integrable
nonequilibrium systems~\cite{Arndt}. The present work is motivated by
an ongoing effort to apply similar techniques to non-integrable
systems such as directed percolation~\cite{Physica}. Our aim is to
understand the nature of {\it nonequilibrium} phase transitions in
more detail and to search for a signature of integrability and
non-integrability in the distribution of Yang-Lee zeros.

To apply the concept of Yang-Lee zeros to DP, we consider the order parameter
in a finite system as a function of the percolation probability $p$ in the
complex plane. This can be done by studying the {\em survival probability}
$P(t)$, which is defined as the probability that a cluster generated in a
single site at time $t=0$ survives up to time $t$ (or even longer). Note that
in the limit of infinite $t$ the order parameter $P(\infty)$ and $P(t)$ coincide. The partition sum
of an equilibrium system and the survival probability of DP show a similar
behaviour in many respects, e.g. they both are positive in the physically
accessible regime and can be expressed as polynomials in finite systems. If
the system size tends to infinity, both functions exhibit a non-analytic
behavior at the phase transition as the Yang-Lee zeros in the complex plane approach the real line.

\begin{figure}
\epsfxsize=95mm
\centerline{\epsffile{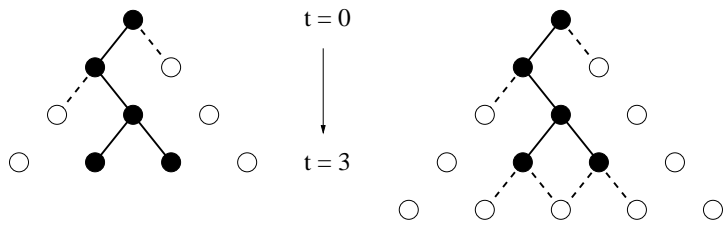}}
\caption{\label{FigCluster}
Example of a cluster which survives until time $t=3$ (solid lines). The
bonds belonging to its hull are shown as dashed lines. Left: The corresponding 
weight to the survival probability $P(3)$ in~(\ref{SurvivalProbability}) is $p^4(1-p)^2$. Right: For the probability $R(3)$ the following row of bonds has to be taken into account as well, contributing an additional factor $(1-p)^4$ thus leading to a different weight in ~(\ref{DieoutProbability}) which is given by $p^4(1-p)^6$.}
\end{figure}

In directed bond percolation the survival probability is given by the sum over
the weights of all possible configurations of bonds. Each conducting bond
contributes to the weight with a factor $p$, while each non-conducting bond
contributes with a factor $1-p$. However, the states of those bonds which do
not touch the actual cluster are irrelevant as they do not contribute to the
survival of the cluster. Therefore, it is sufficient to consider the sum over all possible {\em clusters} {$\cal C$} of bonds connected to the origin. Each cluster is weighted by the contributions of the conducting bonds belonging to the cluster and the non-conducting bonds belonging to its hull. More precisely, the survival probability can be expressed as
\begin{equation}
\label{SurvivalProbability}
P(t) = \sum_{\cal C} p^n (1-p)^m \,,
\end{equation}
where the sum runs over all clusters reaching the horizontal row at time $t$.
For each cluster $n$ denotes the number of its bonds, while $m$ is the
number of bonds belonging to its hull (see left part of
Figure~\ref{FigCluster}). Note that $m$ does not include bonds
connecting sites at time $t$ and $t+1$ since the cluster may survive
even longer. Summing up all weights in
Equation~(\ref{SurvivalProbability}), one obtains a polynomial of
degree $t^2+t$. As can be verified, the first few polynomials are
given by

\begin{eqnarray}
P(0)=1\\
P(1)=2p-p^2 \nonumber \\         
P(2)=4p^2-2p^3-4p^4+4p^5-p^6 \nonumber \\
P(3)=8p^3-4p^4-10p^5-3p^6+18p^7+5p^8-30p^9+24p^{10}-8p^{11}+p^{12} \nonumber \\  P(4)=16p^4-8p^5-24p^6-8p^7+6p^8+84p^9-29p^{10}-62p^{11}-120p^{12} \nonumber \\
\qquad\quad+244p^{13}+75p^{14}-470p^{15}+495p^{16}-268p^{17}+83p^{18}
-14p^{19}+p^{20}\,. \nonumber
\end{eqnarray}  

\begin{figure}
\epsfxsize=100mm
\centerline{\epsffile{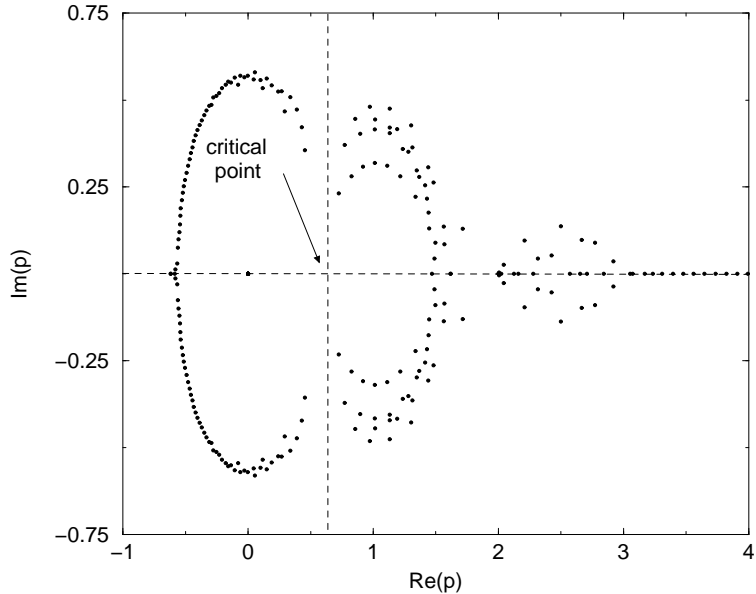}}
\caption{\label{FigYangLee}
Distribution of zeros of the polynomial $P(15)$ in the complex plane.}
\end{figure}

As $t$ increases, the number of cluster configurations grows rapidly,
leading to complicated polynomials with very large coefficients. The
distribution of zeros in the complex plane for the polynomial
$P(t=15)$ is shown in Figure~\ref{FigYangLee}. As can be seen, the
distribution reminds of a fractal, being perhaps a signature of the
non-integrable nature of DP. As expected, the zeros approach the phase transition point from above and below.

While a statistical analysis of the distribution of zeros will be presented
elsewhere, we will focus in the present work on a particularly surprising
observation. This is the existence of certain points on the real line where the polynomials can be solved exactly for all values of $t$. Beside the trivial points $p=0$ (where $P(t)=\delta_{t,0}$) and $p=1$ (where $P(t)=1$), we find a $t$-independent zero at $p=2$ and, even more surprisingly, a very simple solution if $p$ is equal to one of the Golden Ratios
$(1\pm \sqrt{5})/2$. The Golden Ratios are the roots of the quadratic equation
$p^2=p+1$ and play an important role not only in number
theory~\cite{NumberTheory} but also in other fields ranging from chaotic
systems~\cite{chaotic} to arts~\cite{arts}. Although these special points are located outside the physically accessible regime $0 \leq p \leq 1$, their existence may help to understand the structure of Yang-Lee zeros in the complex plane. 


\vspace{5mm}
\noindent
{\bf Time-independent zero at $p=2$:}\\
For $p=2$ and $t\geq 1$ all polynomials $P(t)$ vanish
identically. This can be shown as follows. Let us consider the
probability $R(t)$ that a cluster dies out at time $t$, i.e., the
row at time $t$ is the last row reached by a cluster. Obviously
$R(t)$ is related to the survival probability by
\begin{equation}
\label{RDef}
R(t) = P(t)-P(t+1) \,.
\end{equation}
Clearly, $R(t)$ can be expressed as a weighted sum over the same set
of clusters as in~(\ref{SurvivalProbability}). However, in the present
case the weights differ from those in
Equation~(\ref{SurvivalProbability}) by the number of non-conducting
bonds in the clusters hull between $t$ and $t+1$  since it is now
required that all sites at time t+1 are inactive (see right part of
Figure~\ref{FigCluster}). This means that $R(t)$ can be expressed as
\begin{equation}
\label{DieoutProbability}
R(t) = \sum_{\cal C} p^n (1-p)^m (1-p)^{2k}\,,
\end{equation}
where $n$, $m$ and ${\cal C}$ have the same meaning as
in~(\ref{SurvivalProbability}) and $k$ is the number of active sites in the
horizontal row at time $t$. Obviously, for $p=2$ the additional factor
$(1-p)^{2k}$ drops out so that $P(t)=R(t)$ for all values of $t$. Moreover,
$R(0)=P(0)=1$ (for $p=2$). Combining these results with Equation~(\ref{RDef}) we arrive at $P(t)=0$ for $t>0$, which completes the proof.



\vspace{5mm}
\noindent
{\bf Exact solution for $p$ at the Golden Ratio:}\\
For $p=(1\pm \sqrt{5})/2$ we find that the survival probability `oscillates' between two different values, namely
\begin{equation}
\label{MainResult}
P(t) = 
\cases{
1 & if $t$ is even\\
\frac{\pm \sqrt{5}-1}{2} & if $t$ is odd.\\}
\end{equation}
\begin{figure}
\epsfxsize=120mm
\centerline{\epsffile{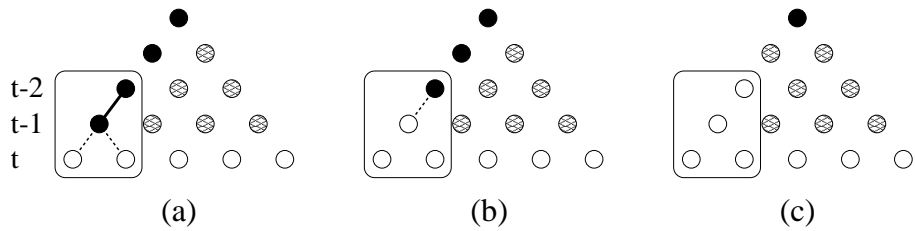}}
\caption{\label{FigProof}
Decomposition of the configurational sum of non-surviving
clusters into three subsets.
Open and closed bonds are denoted by solid (dashed) lines.
Bonds which are not shown may be either open or closed.
The box includes all bonds contributing to the factor $\hat{Q}_1$
while all other bonds contribute to the factor $\hat{Q}_2$. The
proof shows that the configurations in (a) and (b) cancel
so that only the configurations of (c) contribute to $Q$. Iterating
the procedure by shifting the box to the right, it can be
shown that all sites at $t-2$ and $t-1$ have to be zero.}
\end{figure}

To prove this result, we first verify that~(\ref{MainResult}) is indeed
satisfied for $t=0$ and $t=1$. Then we show that 
\begin{equation}
P(t) = P(t-2) \qquad \mbox{for } t\geq 2 \qquad \mbox{and}\qquad p=(1\pm \sqrt{5})/2\,.
\end{equation}
However, instead of analyzing the survival probability directly, it
turns out to be more convenient to consider the complementary probability
$Q(t)=1-P(t)$ that a cluster does {\em not} survive until time
$t$. Obviously, $Q(t)$ is the sum over the weights of all clusters
which do not reach the horizontal row at time $t$, i.e., we impose the
boundary condition $s_{0,t}=s_{1,t}=\ldots=s_{t,t}=0$. Depending on
the states of the two sites $s_{0,t-1}$ and $s_{0,t-2}$ at the left
edge of the clusters, this set of clusters may be separated into three
different subsets, namely,\\

        {\bf(a)} a subset where $s_{0,t-1}=s_{0,t-2}=1$,

        {\bf(b)} a subset where $s_{0,t-1}=0$ and $s_{0,t-2}=1$, and

        {\bf(c)} a subset where $s_{0,t-1}=s_{0,t-2}=0$.\\

\noindent
Next we show that the weights of the clusters in the subsets (a) and (b) cancel each
other. To this end we note that the weighted sum $\hat{Q}(t)$ over all
clusters in both subsets may be decomposed into two independent factors
$\hat{Q}(t)=\hat{Q}_1\hat{Q}_2$, where $\hat{Q}_1$ depends only on the state of the three bonds
between the sites $s_{0,t-2}$, $s_{0,t-1}$, $s_{0,t}$, and $s_{1,t}$ (inside
the box in Figure~\ref{FigProof}), while $\hat{Q}_2$ accounts for all other relevant
bonds. Obviously, the first factor is given by
\begin{equation}
\hat{Q}_1^{(a)} = p(1-p)^2 \,, \qquad \hat{Q}_1^{(b)} = 1-p \,,
\end{equation}
while $\hat{Q}_2$ takes the same value in both subsets. Thus, if $p$ is
given by the Golden Ratio, we obtain $\hat{Q}_1^{(a)}+\hat{Q}_1^{(b)}=\hat{Q}_1(t)=0$
and therefore the weights of subsets (a) and (b) cancel. Consequently,
all remaining contributions to $Q(t)$ come from the clusters in subset
(c) where the sites $s_{0,t-1}$ and $s_{0,t-2}$ are inactive.

Now we can iterate this procedure by successively considering the
sites $s_{j,t-1}$ and $s_{j,t-2}$ from the left to the right, where
$j=1\ldots t-2$. In this way it can be shown  that all these sites
have to be inactive as well. Therefore, the only surviving
contributions are those in which the entire row of sites at $t-2$ is
inactive, implying that $Q(t)=Q(t-2)$. The proof of
Equation~(\ref{MainResult}) then follows by induction.

To summarize, we have shown that the survival probability $P(t)$ of a
(1+1)-dimensional directed bond percolation process can be computed
exactly at certain points on the physically non-accessible part of the
real axes. We hope that this observation may help to understand the
distribution of Yang-Lee zeros in the complex plane. Moreover,
we expect similar special points to exist in other realizations of DP
and related models.
\ack
This work was supported by the DAAD/CAPES within the German-Brazilian
cooperation project PROBRAL -- ``Rigorous Results in Nonequilibrium
Statistical Mechanics and Nonlinear Physics''.


\end{document}